# Laser pulse shape designer for direct-drive inertial confinement fusion

Tao Tao (陶弢)[1*], Guannan Zheng(郑冠男)[1], Qing Jia(贾青)[1,3], Rui Yan(闫锐)[2,3], Jian Zheng (郑坚)[1,3†]

[1]Department of Plasma Physics and Fusion Engineering and CAS Key Laboratory of Geospace Environment, University of Science and Technology of China, Hefei, Anhui 230026, People's Republic of China
[2]Department of Modern Mechanics, University of Science and Technology of China, Hefei, Anhui 230026, People's Republic of China
[3]Collaborative Innovation Center of IFSA, Shanghai Jiao Tong University, Shanghai 200240, People's Republic of China

*tt397396@ustc.edu.cn
†jzheng@ustc.edu.cn

We have developed a pulse shape designer for direct drive inertial confinement fusion with the goal of achieving high compression while keeping hydrodynamic instability at a tolerable level. The laser and compression shockwaves imprint unique perturbation patterns on the target that serve as seeds for later instability development, and only after modeling these cascaded processes can the design be fully optimized. The designer scans pulses using linear instability analysis on one-dimensional implosion simulations and builds a predictive surrogate that relates pulse shape to implosion performance. Clustering methods are used to select a small number of key pulse shapes for two-dimensional evaluation, extract the imprint seed information, and use it to calibrate the surrogate model. This prediction-correction alternation produces the optimal pulse shape taking into account the facility-specific laser speckle and target pellet roughness characteristics. When tested on the double-cone ignition scheme [J. Zhang, Phil. Trans. R. Soc. A. 378.2184 (2020)], the optimized pulses achieve an increase in areal density, showing a significant capability in short to medium wavelength mitigation. The resistance of the pulses to laser shaping errors is also elucidated. The pulse shape designer balances search speed and prediction credibility, which is expected to improve fuel quality in direct drives and reduce experimental risk.

## 1. Introduction

Inertial confinement fusion (ICF) [1,2] uses a driver such as a laser to implode a pellet target, compressing its fuel shell to extreme conditions[3] and triggering a sustained thermonuclear reaction. Self-heating of alpha particles requires an areal density $\rho_a > 0.3 \text{g/cm}^2$ of the final assembly. However, inwardly accelerated shells create favorable conditions for the development of Rayleigh-Taylor Instability (RTI), which can lead to severe ablator-fuel mixing and even shell breakup[4,5]. Direct-drive ICFs[6] are particularly susceptible to RTI because laser imprinting[6,7] can directly couple laser illumination non-uniformities to the shell, providing a large amount of seeding for RTI. Methods to suppress seeding include pellet roughness reduction[8], beam overlap optimization[9,10], and beam smoothing[6].

Pulse shaping[11,12] is another powerful and easy-to-implement tool that is useful for both seed mitigation and in-flight stability control. Pickets are added to the temporal waveform of the pulse, which optimizes the scale length of the plasma and increases the mass ablation rate and thermal smoothing intensity. The pulse waveform is most often timed to meet the requirements of "adiabat shaping"[13,14]: the outer layer of the capsule is heated by an intense but unsupported shock[15] to obtain a high imprinting resistance, while the inner layer undergoes a nearly isotropic compression[16] that keeps its entropy low to maintain compressibility.

The nonlinear nature of implosions makes their performance highly sensitive to the laser target initial conditions. Metrics such as the Ignition Threshold Factor (ITF)[17,18] are available to help specify the ignition margin, which is usually summarized from historical data into an explicit algebraic expression. For laser and target pellet perturbations with stochastic properties, the statistical description of the ITF is often insufficient. Fine-tuning optimization based on integrated experiments or computer simulations can adjust the pulse shape to account for the laser and target pellet characteristics of a particular device. Unfortunately, integrated experiments are extremely expensive, and direct numerical simulations are beyond the capabilities of current computational hardware. Iterations of implosion schemes necessarily rely on reduced models.

Recently, data-driven approaches have shown great potential in implosion optimization[19]. The paradigm consists of design, regression/classification, and calibration. On the design side, the maximum yield can be found by automatically adjusting the simulation inputs using evolutionary schemes such as genetic algorithms[20-22]; the surrogate model is based on regression/classification on a large number of implosion experiments and simulations. Algorithms such as Gaussian process regression and random forest can identify high-dimensional correlations between experimental inputs and outputs[20,23,24]. Regression schemes achieve improved experimental neutron yields without actual physical modeling[25]. Calibration uses sparse experimental points to update the surrogate model originally built on a large amount of synthetic data, increase its credibility. A typical example is transfer learning[26, 27].

We have developed an automated pulse shape designer. The designer obtains large amount of data in one dimensional pulse simulations. Post-processing evaluates each pulse shape based on fuel compressibility and stability, and then uses classification to build a predictor for pulse performance. A small number of high-dimensional, high-resolution simulations were then used to calibrate this predictor. The calibration pulses are automatically selected by a clustering method. This novel prediction-correction approach allows the surrogate model to account for both the pulse data volume and detailed imprint signatures. The designer also includes robustness evaluation based on Hessian matrix analysis, the pulses' resistance to realistic laser power and synchronization errors are clarified, the most severe error patterns are identified. The designer has been used for a novel Double Cone Ignition (DCI)[28] scheme considering its specific laser-target characters. The results show that the optimized pulses are expected to improve the existing areal density and stability.

The chapters are organized as follows. Section 2 introduces the optimization workflow and some fundamentals of DCI; Section 3 introduces the linear growth prediction of implosion RTI; Section 4 introduces the implosion performance surrogate model; Section 5 introduces the model correction by imprinting seeds; Section 6 introduces the pulse quality check; Section 7 is a summary.

## 2. Optimization workflow

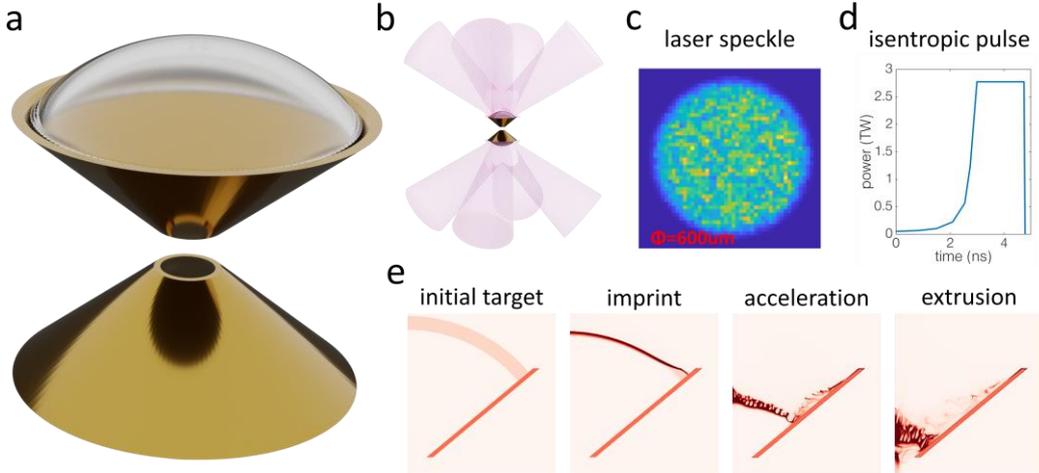

Fig. 1: (a) Schematic of the DCI target; (b) 4x2 incident laser configuration; (c) simulated laser speckle; (d) isentropic drive pulse used in the example simulation; (e) shell density for four implosion stages: initial target, imprint, acceleration, and extrusion.

DCI uses a head-on pair of cones to guide the embedded fuel cap, as shown in Fig. 1. In the DCI scheme, multiple laser beams overlap on one side of the cone, compressing and accelerating the shell, the fuel is extruded from the smaller hole of the cone, and the oppositely traveling fuel packages collide and stagnate, increasing density and pressure to form an assembly ready to be ignited by relativistic electron beams (fast ignition). The main advantage of the DCI over conventional spherical targets is the smaller illuminated solid angle, and the overall construction cost will be lower if the high drive energy requirement is relaxed.

However, the characteristics of DCI also pose the challenge of implosion instability: the direct drive makes it easier to couple beam speckles to the fuel, and the short-wavelength non-uniformity of the laser can eventually seed the rapid development of RTI, threatening the integrity of the shell. The presence of cone boundaries can also break the symmetry of spherical illumination and increase long-wavelength perturbations. The risk of instability is illustrated by the isentropic drive simulation shown in Fig. 1(e), with laser and target conditions taken from recent DCI experiments: The laser power delivered to each cone is 6 kJ, the four lasers are uniformly distributed around the central axis of the cone at the same angle of 50° to the axis. The cone opening angle is 100°. The small cone hole radius is 50 μm and the double cone separation is 100 μm, the cone itself is fixed in the simulation because we only study the fuel instability at this stage; the fuel composition is 1.27g/cm$^3$ of high-density plastic (C:H=1:1) and the fuel is a cap with an inner radius of $r_i$= 450 μm and an outer radius of $r_o$= 490 μm. It can be seen that the isentropic pulse does not reach its predicted compression in the 1D simulations, the shell experiences severe RTI at the outer surface and breaks up before leaving the cone. The combination of beam speckle and overlapping asymmetry perturbs the shell during the imprinting phase. Measures must be taken to increase the stability of the shell.

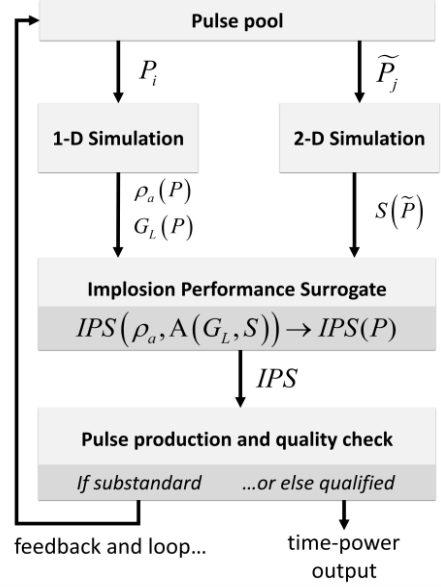

Fig. 2. Pulse shape designer workflow, with the primary goal of maximizing the value of the implosion performance surrogate IPS.

Finding the optimal pulse is not an easy task, because pulse shapes can vary continuously, the pulse configuration space is extremely large, and the pulses that satisfy the adiabat shaping and areal density requirements occupy only a small percentage of the configuration space. To fully interpret the simulation results while improving optimization efficiency, the designer uses a machine learning approach to construct an Implosion Performance Surrogate (IPS), this IPS is also known as a fitness function in machine learning jargon. The IPS function takes the maximum areal density $\rho_a$ and the maximum perturbation amplitude A as inputs, and its algebraic expression is given in Chapter 4. In turn, $\rho_a$ and A implicitly depend on the pulse shape P through the specific implosion process, and the goal of optimization is to adjust P so that IPS is maximized.

The optimization process shown in Fig. 2, consists of four main components: the first component, a one-dimensional (1-D) simulation branch, uses the 1-D Lagrangian hydrodynamics code MULTI[29] to obtain the fluid state in the compression direction, and takes the algebraic scaling law[30] to estimate the growth rate $G_L$ of RTI in the acceleration phase. It features fast simulation speed and is the primary data source for the surrogate model. The second component, a 2-D simulation branch, uses the multidimensional Eulerian radiation hydrodynamics code FLASH[31] on key pulses, with realistic device beam overlap and speckle patterns. It models nonlinear processes such as imprinting and Richtmyer-Meshkov Instability (RMI), packing laser and seed relations into an interpolation function S. A third component, the construction and updating of the implosion performance surrogate IPS, which incorporates linear RTI growth prediction and nonlinear imprint seed correction. The fourth component, pulse production and quality check. Operate the surrogate to determine a best pulse, put it into a 2-D testing for its actual compressibility and stability. If the pulse product meets the experiment standard, output and terminate, otherwise use the surrogate to determine a new batch of sampling points, put them into the pulse pool and loop the whole process. The key here is the prediction-correction alternation, where after several loops the surrogate model fully samples the pulse configuration space and can also take into account the device-specific imprinting information. The optimization algorithm is implemented using the Matlab language and its machine learning toolkit.

## 3. RTI linear growth prediction

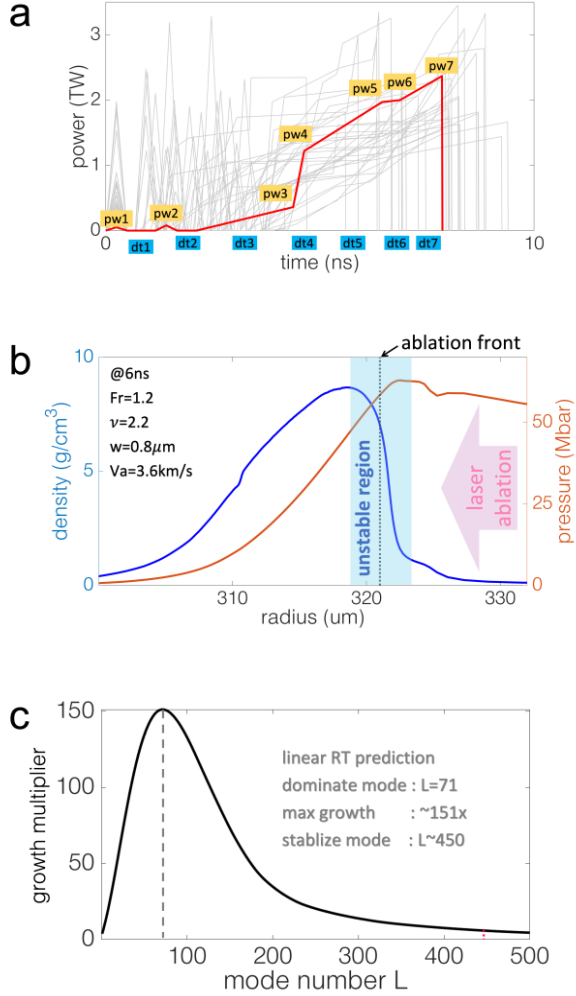

Fig. 3: (a) Series of two-picket pulses with 14 degrees of freedom; (b) fitting of hydrodynamic variables near the ablation front to obtain key parameters related to the linear RTI growth rate; (c) calculated RTI growth multiplier for the red example pulse above.

The designer uses a semi-analytical model to estimate the linear growth rate of the implosion. First, the continuous pulse profile is decomposed into finite degrees of freedom (DoF). Fig. 3(a) shows an example two-picket pulse series, both starting with two 500-ps delta waves followed by a monotonically rising 5-free-node main pulse. dt is the variable time interval, pw is the variable power level, and the example pulses have a total of 14 DoFs, i.e., the configuration space is 14-dimensional. In the second step, 1-D implosion simulations are performed, and the ablation front is detected at each evolution step, and the region near it with opposite density and pressure gradients is delineated as RTI unstable, an example of which is shown in Fig. 3(b). The instantaneous growth rate γ was calibrated using the steady-state growth rate from the work of Betti[30]:

$$\gamma = \sqrt{\alpha_1(\nu,w)kg + \alpha_2(\nu,w,F_r)k^2V_a^2} - \beta_1(\nu,w)kV_a \qquad (Eq.1)$$

Where k is the wave vector, g is the effective acceleration, $V_a$ is the ablation velocity, $\nu$ is the power index of heat conduction ($\nu = 2.5$ for electron conduction), w is the characteristic width of the ablation front, $F_r = V_a^2/gw$ is the Froude number, and $\alpha_1, \alpha_2, \beta_1$ are complex functions that can be found in Betti's work. The values of $\nu$, $w$, $F_r$ and $V_a$ can be determined by fitting on hydrodynamic observables such as density and pressure in the unstable region. The annotation in Fig. 3(b) shows the fitted values at +6 ns after the laser is turned on. In our practice, the value of g is calculated by tracking the shell center of mass trajectory. The third step is to integrate over all simulated time slices 1 to n to obtain the growth multiplier $G_L = \exp(\gamma_{1L} \Delta t_1 + \gamma_{2L} \Delta t_2 + \cdots \gamma_{nL} \Delta t_n)$ for each circular mode number $L = kr$, r is the radius of the shell. Multiplying the growth rate spectrum $G_L$ and the seed spectrum S, we can obtain the perturbation amplitude spectrum A. In Fig. 3(c), one can identify the peak RTI growth mode and the stabilization mode corresponding to this pulse shape.

An important assumption for estimating RTI growth using algebraic scaling is steady-state ablation, i.e., the fluid distribution reaches quasi-equilibrium. Most of the RTI growth occurs during the sustained acceleration phase driven by the main pulse, when the plasma corona region is largely established, and the fluid is able to respond to the pulse power change. As for the picket and the beginning of the main pulse, although the steady-state ablation assumption is not valid, the laser power is lower at this time. The fuel center of mass acceleration is small, and its contribution to the RTI is also small, resulting in negligible error. Overall, this time integral instability estimation method is generally acceptable.

## 4. Implosion performance surrogate and pulse shape evolution

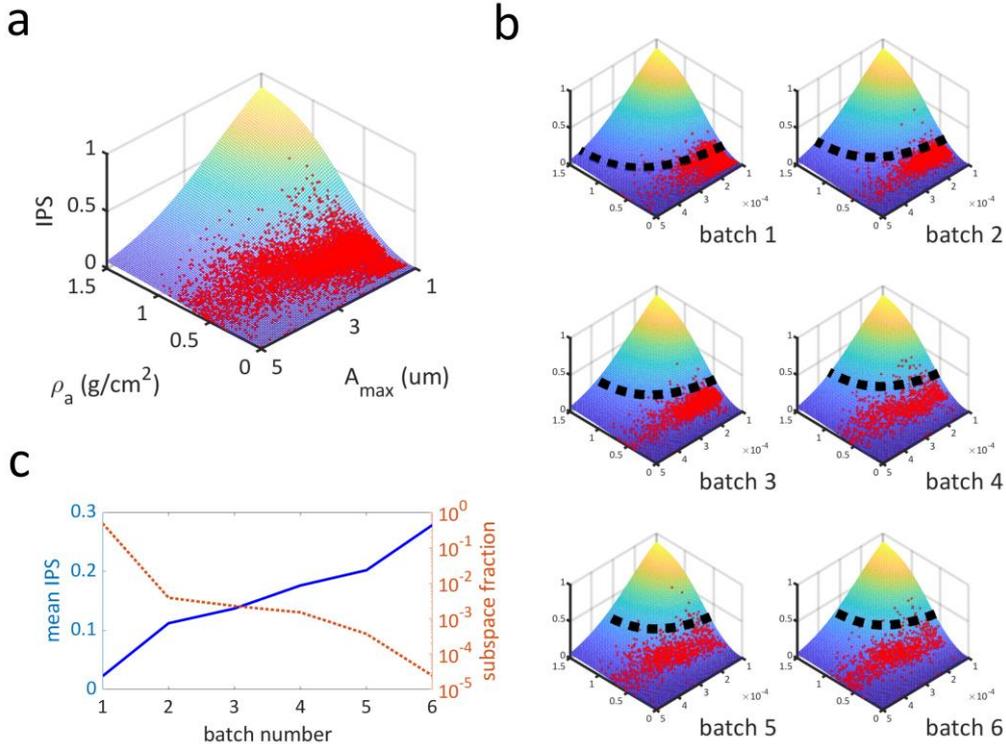

Fig. 4: (a) IPS mesh shape depending on the stagnation areal density $\rho_a$ and the maximum perturbation amplitude A. One red dot corresponds to one pulse sample; (b) 6 batches of pulse optimization, the black dashed line marks the thresholds for true/false

classification, the values of the 6 thresholds are 0.1, 0.15, 0.2, 0.25, 0.275, 0.3; (c) The left axis shows the mean IPS score, the right axis shows the proportion of "islands of high performance" to the total configuration space volume.

The designer needs to quantitatively evaluates the implosion quality. Based on our current knowledge of DCI physics, the IPS is explicitly expressed as:

$$\text{IPS}(\rho_a, A) = [1 - 2/(1 + \exp(\rho_a/0.7 \text{g} \cdot \text{cm}^{-2}))^2] \cdot \exp(-(A/3\mu m)^2) \quad (Eq.2)$$

The square bracket term on the right side of Eq. 2 represents the compressibility, and the exponential term on the far right represents the stability. Both terms are normalized so that the IPS value is also bounded between 0 and 1, the higher the better.

The two scaling factors $0.7 \text{ g/cm}^2$ and $3\mu m$ in Eq. 2 are heuristic values based on the current physical understanding of DCI: alpha particle self-heating requires $\rho_a > 0.3 \text{g/cm}^2$, corresponding to a basic compression term score ~0.4; ignition electron beam deposition requires $0.8 \text{ g/cm}^2$ [28], corresponding to an excellent compression term score ~0.8. The isentropic pulse imprint amplitude is typically ~3 μm, if the ablation stabilization is sufficient in the acceleration phase, the perturbation will no longer be amplified, so 3 μm is a characteristic value, corresponding to a stability term score ~0.4; if the picket can be further optimized, the imprint can be cut in half to ~1.5 μm, the stability term can be given an excellent score ~0.8. The optimal pulse shape is a trade-off between compressibility and stability, and the IPS expression is formulated by considering these factors.

Beginning to build the surrogate model. Training data is obtained by sampling multiple batches of pulses. 1-D simulations are carried out to obtain pulse $\rho_a$ and A, then the IPS is calculated from Eq.2. Each batch is pre-specified with an IPS threshold, and the points above/below the threshold are assigned the labels T(true)/F(false), respectively, to form a labeled training set for the classification problem. The classification algorithm we use is a Gaussian kernel Support Vector Machine (SVM). Regression is not used here, because our tests show that regression has a larger variance in our sparsely sampled situation, while classification performs better at following contours, and is more capable at identifying "islands of high-performance pulses" in the configuration space.

If the optimal pulses still do not meet the standard, the classification algorithm is used to filter a new batch of pulses and send them to the next cycle. Any random pulses predicted by the classifier to be below the threshold are discarded and do not enter the hydrodynamic simulation, and the positively predicted pulses are cached in the pulse pool. Pulse performance increases with the filter threshold, and the boundaries of the "islands of high-performance pulses" becomes clear. At this point, the optimal pulse shape can be determined by the best from all the batches, or by using a generalized implicit function search algorithm such as Nelder-Mead (fminsearch in Matlab).

Fig. 4(b) shows six batches, each containing 2000 pulses, Fig. 4(c) shows that the batch-averaged IPS score increases from ~0.02 to ~0.3, and identifies the high-performance subspace above the threshold, which accounts for only $2 \times 10^{-5}$ of the total configuration space volume. The top 10% sampled pulses can achieve a compression $\rho_a > 0.5 \text{ g/cm}^2$ with a maximum imprinted amplitude $A < 2 \mu m$, demonstrating the effectiveness of the algorithm.

It needs to be clarified why pulse shape evolution does not use the common evolutionary algorithm paradigm, for two reasons: First, while the IPS can be expressed explicitly in terms of $\rho_a$ and A, A depends on the imprint seed S, and S is corrected (detail in the next section) in each loop. As the prediction-correction alternates, the extrema of the optimization problem actually change. Using a genetic algorithm, for example, helps converge quickly to the extrema implied by the known pulse data, but it hardly helps in the next batch run after correction. Instead, the current algorithm is able to disperse points more evenly in the configuration space, which is more helpful in the subsequent search for pulse aggregation features. Second, batch simulations facilitate parallelism and are less difficult to schedule, allowing us to run a large number of 1-D implosion simulations simultaneously.

## 5. Model correction by imprinting seeds

A major goal of pulse shaping is to limit the amplitude of the dominant mode to a tolerable level, and the shaping strategy must be determined in conjunction with the imprint seeds. We tuned the FLASH fluid program to reflect the real DCI perturbation signature: target roughness $\Delta_{std}$ = 45 nm, laser non-uniformity $\sigma_{rms}$ ~ 15%, multi-beam overlap pattern included. By analyzing the center-of-mass perturbation of the fuel, the seed spectral corresponding to circular modes L is obtained. The FLASH simulation is performed in 2-D cylindrical geometry.

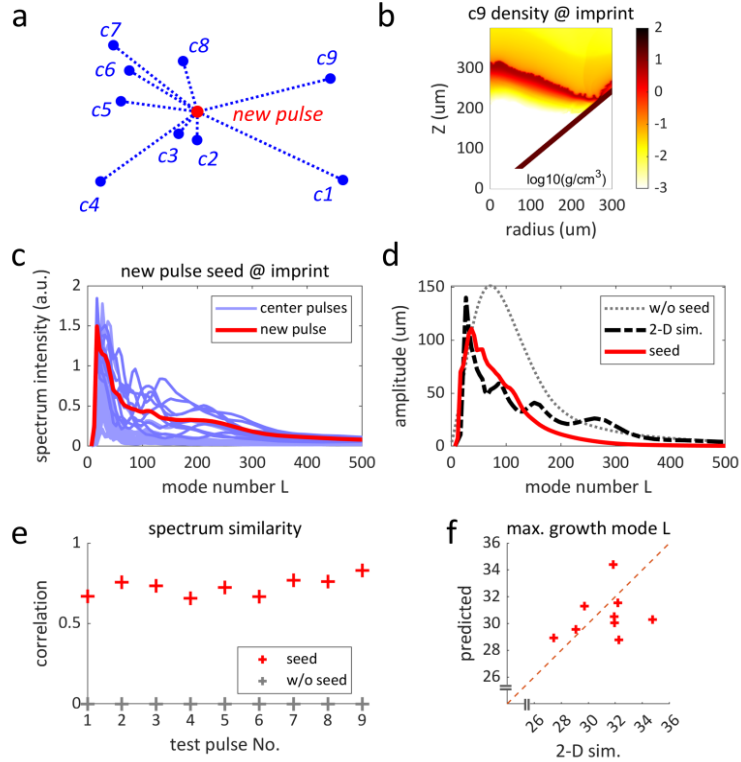

Fig. 5: (a) Illustration of clustering, distances of a new pulse from the center pulses are indicated by the blue lines; (b) C9 pulse density at imprint; (c) Interpolation synthesis of new pulse seeds; (d) Comparison of amplitude predictions with and without seed correction, using 2-D simulations as facts; (e) Correlation between predicted spectrum and 2-D simulated fact spectrum; (f) Dominant mode comparison between seed-corrected prediction and 2-D fact.

It is not practical to examine all pulse samples in 2-D, so we use a clustering method to determine several key pulses, extract their seed features, and use a weighting method to synthesize seeds for all newly sampled pulses. Clustering is achieved by minimizing the sum of high-dimensional Euclidean distances between all points and allows grouping of the unlabeled data. In each batch, high quality pulses form clusters in the configuration space, and the K-means algorithm is used to divide them into K separate groups and thus find the "center of mass" coordinates of each group. Currently, K is set to 9, and the centers c1-c9 are shown in Fig. 5(a). 6 batches need 5 rounds of feedback, resulting in a total of 45 centers. The density of the flying shell at 2/3 of the initial radius is analyzed and the imprint spectrum is obtained using wavelet analysis. In the next batch, the seed $S(\widetilde{P})$ of the newly sampled pulse can be obtained by interpolation:

$$S(\widetilde{P}) = \sum_k S_k \frac{||\widetilde{P}-c_k||^{-2}}{\sum_j ||\widetilde{P}-c_j||^{-2}} \quad \text{(Eq.3)}$$

Where $S_k$ is the known central pulse spectrum, the right-side fraction is the summation weight, c is the central pulses' coordinates. The weights decrease inversely, as the distance between the new pulse $\widetilde{P}$ and the central pulse c increases. Fig. 5(c) shows the synthesized seed for the isentropic pulse.

The prediction of the instability changes significantly after the correction. Fig. 5(d) shows the prediction of the dominant mode shifted from L=71 without correction to L=37 with correction, which is much closer to the 2-D simulation fact.

There are 9 newly generated pulses that served as a validation set for spectrum synthesis. Fig. 5(e)-(f) show that the predicted spectrum of the corrected surrogate has high correlation (treat the two spectra as signal trains) with the simulation facts, confirming the fidelity of the correction method.

There is a lot of established work in ICF research that uses a small amount of data to calibrate the surrogate, such as transfer learning[26]. In contrast, our designers use an algebraic approach to coupling imprinting and RTI linear growth, which may lose some of the generalizability. However, this has the advantage of fewer hyper-parameters and a clearer physical meaning of the cascaded instability development.

**6. Pulse quality check**

This section performs a quality check on the optimized pulse products, evaluating them in three aspects. The first is the improvement over the currently used experimental pulse, the second is the robustness against shaping errors, and the third is a comparison of the optimization with and without corrections.

6.1. Progress relative to existing experimental pulses

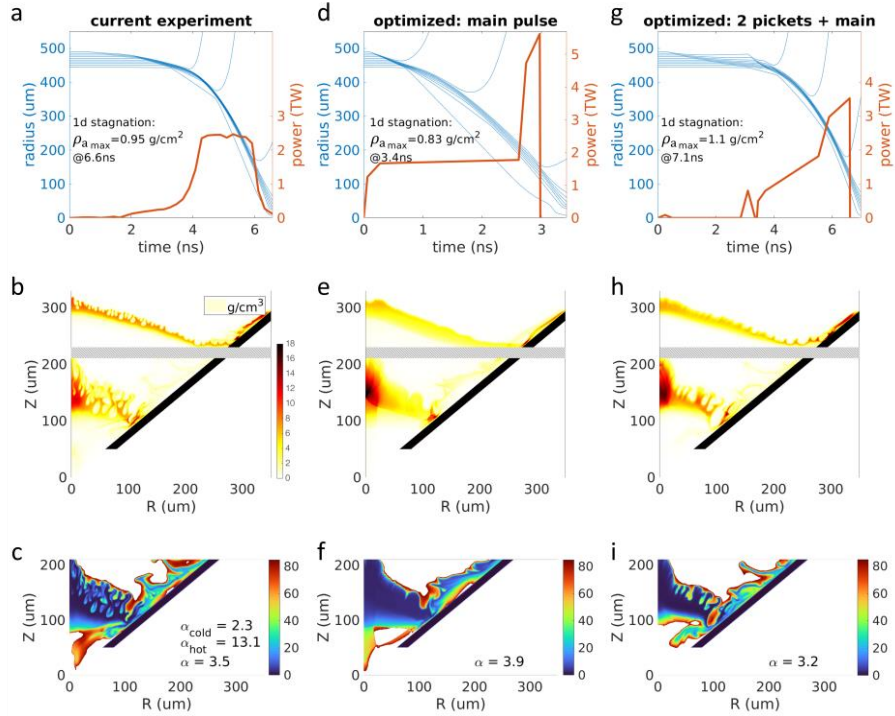

Fig. 6. (a-c) Implosion simulation of the current experimental pulse; (d-f) Implosion simulation of the designer-optimized acceleration pulse; (g-i) Implosion simulation of the designer-optimized two-picket pulse. The first row shows the respective pulse shapes and 1-D implosion streamlines. The second row shows the shell density at 2/3 and 1/3 of the initial radius. The third row is the shell adiabat at 1/3 of the initial radius.

The currently used pulse shape for the DCI experiment is shown in Fig. 6(a), which is a high compression design with a 1-D areal density expectation of $\rho_a = 0.95$ g/cm$^2$. At 2-3.5 ns, the pulse drives a supported shock into the shell and achieves a relatively high IFAR ~20 (In-Flight Aspect Ratio, defined as the compressed fuel thickness divided by the initial radius). The perturbation pattern on the outer surface can be seen in Fig. 6(b), where the plasma corona scale length is too short to smooth the dominant wavelength at $\lambda \sim 10$ μm. The seeds are further amplified and evolve into the nonlinear phase at the end of the acceleration, where the bubbles are seen to reach the inner surface of the shell, with L ~ 45 being the dominant mode. The average adiabat of the cold fuel (sampling point along the shell mass center) is kept low at $\alpha = 2.3$, but the fingers with high adiabatic $\alpha = 13.1$ penetrate deep and contaminate the inner fuel. The hydrodynamic efficiency (the fraction of laser energy converted to shell energy) of this quasi-isotropic pulse is 9.2%.

After 12 batch loops and 24,000 test cases, the designer provided an optimal acceleration pulse as shown in Fig. 6(d): The laser power rises rapidly to 1.8 TW before the first shock reaches the inner surface, and then remains steady for 2.2 ns, during which time the shell is not further compressed. The pulse power rises to 5 TW at the very end, producing a final boost similar to the shock ignition. The imprinted seeds are neutralized by strong mass ablation, the accelerated RTI is mitigated by a prolonged hot corona. The shell remains intact and there are no significant short or medium wavelength perturbations in the density or adiabat. For this pulse it achieves $\rho_a = 0.83$ g/cm$^2$, IFAR ~10, and a slightly high average adiabat value, $\alpha = 3.9$. This pulse sacrifices compression for stability. The hydrodynamic efficiency of this pulse is 8.4%.

The designer performs best when pickets are allowed, as in Fig. 6(g): a low-intensity picket first produces a thin corona, a second, higher picket sends a relatively strong shock, and the main pulse quickly follows. These three shocks reach the inner layer synchronously. IFAR~15, 1-D predicts $\rho_a = 1.1 \text{ g/cm}^2$, this value is the best among all three pulses. More importantly, the outer layer ripple is much less developed due to a well-established plasma corona before 3.5 ns. The bubbles developed at the final acceleration but did not penetrate the homogeneous inner fuel. In Fig. 6(i), the α = 3.2 cold inner layer fuel remains sufficiently thick, indicating good adiabat shaping. The velocity of the shell before exiting the cone is V ~ 190 km/s with a hydrodynamic efficiency of 10.8%.

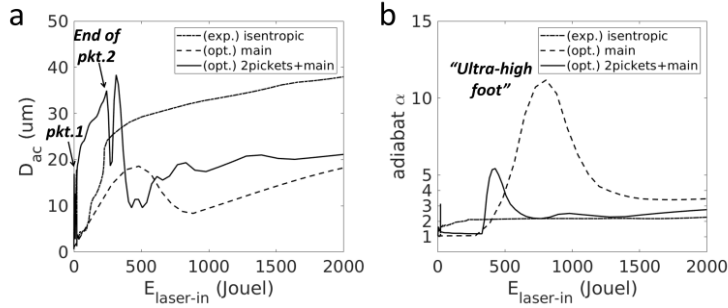

Fig. 7. (a) Evolution of $D_{ac}$, the distance between the ablation front and the critical surface. The horizontal axis is the laser energy delivered, only the initial 2 kJ is shown; (b) Evolution of the fuel adiabat α.

The solid line in Fig. 7(a) reflects the advantageous strategy of the two-picket pulse: the first picket delivers <1% of the total laser energy, the rarefaction up to ~2 ns makes it possible to form a corona with $D_{ac} > 20\mu m$, which is comparable to a typical laser speckle on scale, the illumination non-uniformity decays exponentially in the long-distance traversing from the critical surface to the ablation front. The second picket relays the compression in an imprint safe environment. In contrast, both the experimental pulse and the optimized main pulse lack plasma buffer formation process.

Fig. 7(b) confirms the necessity of a "high foot". For two-picket pulse, the first picket produces an unsupported decaying shock with minimal fuel heating. The second picket creates a slightly higher foot of α~5 for the purpose of ablation stabilization. The adiabat then returns to α~2.5, similar to an isentropic pulse. For the acceleration-only pulse, the already strong imprint seed forced it to adopt a stronger shock, creating an α~12 "ultra-high foot" that preserves the integrity of the shell through excessive amounts of thermal smoothing, but at the expense of areal density. Overall, Fig. 7 demonstrates the designer's adiabat shaping capabilities.

6.2. Robustness of the optimization against shaping errors

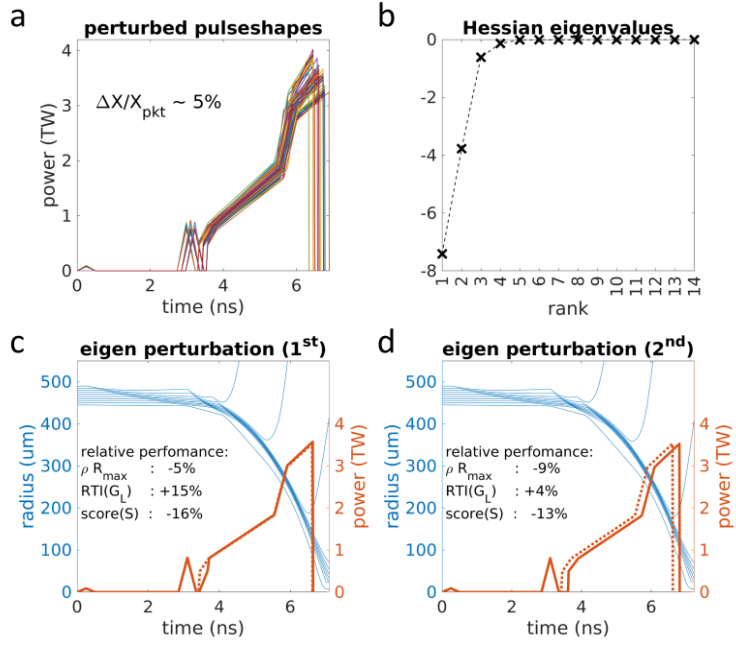

Fig. 8. (a) Adding a 5% perturbation to the optimal pulse; (b) Eigenvalues of the Hessian matrix; (c) Pulse shape and implosion streamline after perturbation along the first eigenvector and (d) along the second eigenvector. The dashed lines are the unperturbed pulse shape.

Real-world lasers inevitably contain shaping errors, and the effect of such errors on optimization results can be assessed by Hessian matrix analysis. The Hessian is the Jacobian of the gradient of a scalar function $H = J(\nabla IPS)$. H describes the local curvature of the multivariable function IPS, and its eigenvalues and eigenvectors reveal the sensitivity and robustness to perturbations. Taking the two-picket pulse with 14 degrees of freedom as an example, the resulting H is a square matrix of rank 14. Fig. 8(b) shows that this matrix is negative definite, proving that the designer has indeed found a local maximum for the IPS score and the optimization has reached the convergence condition.

The first two Hessian eigenvalues are significant, their corresponding eigenvectors indicate the direction of the main descent gradient of the IPS. From the eigenvectors: the primary error is the rising slope of the main pulse at start-up, results in 15% increase in RTI. The secondary error is the time synchronization between the picket and the main pulse, result in 9% decrease in areal density. This reminds us to pay special attention to such kind of shaping errors in the real world.

6.3. Optimization with and without seed correction

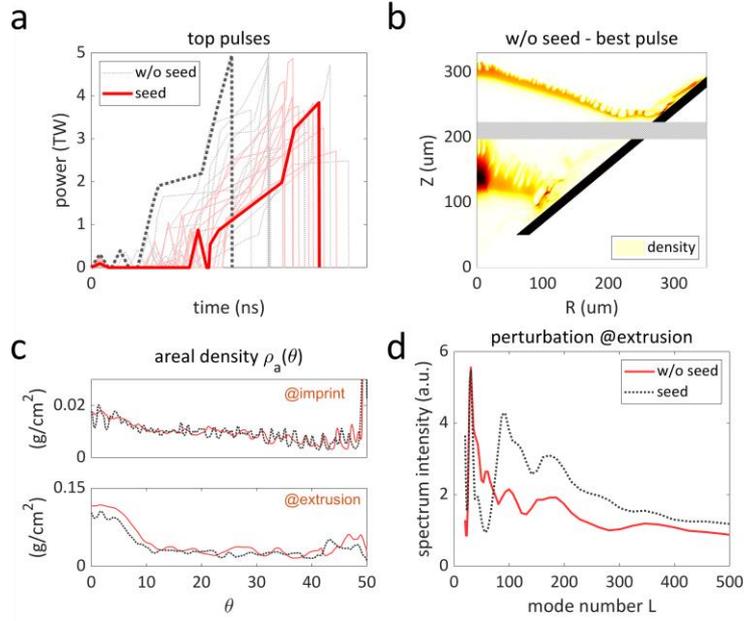

Fig. 9. Comparison of the two optimized series, the red solid line is the optimized series with seed correction and the black dashed line is the optimized series without seed correction. (a) Background shows the top 10 pulse shape of each series, and the two foreground lines show their best; (b) imploding shell density with the best pulse in uncorrected series, snapshot taken at 1/3 and 2/3 of its initial radius; (c) areal density distribution on polar angles of the two best pulses, taken at imprint and extrusion moments; (d) comparison of the perturbation spectrum at the moment of extrusion.

Fig. 9 shows the comparison of the optimization results with and without the seed correction. When the designer sets the seed spectrum to be uniformly distributed, the optimization tends to produce a more temporally compact pulse sequence. From Fig. 9(a) it can be seen that the uncorrected series has a shorter power peaking time and a smaller pulse structure interval. In particular, the interval between two pickets is significantly shorter in the uncorrected series. The uncorrected series underestimates the short-wavelength perturbations and results in insufficient imprint suppression. This conclusion is supported by Fig. 9(c). In the extrusion moment, Fig. 9(d) shows the uncorrected pulse has a relatively weaker stabilization for L>70, means the instability fingers and bubbles are more developed and the mixing risk is higher.

The dominant instability modes of both series are around L~45, such a stable presence of long wavelength perturbation is the combined result of the finite number of beams and the cone boundary, and is difficult to eliminate by pulse shaping alone. Our group is also working on the mid-Z-doped[32] and foam-coated targets[33] for DCI, hoping to conduct joint laser-target optimization in the future.

### 7. Summary

We investigate an automated pulse shape design scheme to improve the integrated performance of ICF direct drive implosions. It forms a primitive prediction of the fuel compressibility and instability through a large number of dimensionally reduced simulations, and incorporates advanced correction on imprint seeds through several key pulse simulations. Classification, clustering, and other algorithm are used to encode implosion performance in a machine learning surrogate. The prediction-correction loop drives the evolution of the surrogate model, generating high-performance pulse shapes, these pulse shapes have good effects on suppressing short to medium wavelength perturbations.

The imprint seed correction is a distinctive feature that allows optimization to reflect the laser non-uniformity and target roughness character of a specific device, making the pulse shaping more relevant and efficient, full implosion simulations verified the necessity of such correction. With its remarkable adiabatic shaping capability, the optimized pulse product is expected to reduce the imprint impact to the current experiment, and increase the assembling quality of the stagnated fuel.

*Acknowledgements*
*This work is supported by the Strategic Priority Research Program of the Chinese Academy of Sciences, Grant No. XDA25010200, the Fundamental Research Funds for the Central Universities, Grant No. WK2140000014, and DCI joint team.*